\documentclass[12pt]{article}
\begin{document} 

\vspace{0.5cm}

\begin{center} 

{\bf{CLASSICAL DEFOCUSSING OF WORLD LINES IN HIGHER DIMENSIONS}} 

\vspace{0.5cm}

R. Parthasarathy {\footnote{sarathy@cmi.ac.in}} \\
The Chennai Mathematical Institute, \\
H1, SIPCOT IT Park, Siruseri, Chennai 603103, India. \\

\vspace{0.5cm}

K.S. Viswanathan {\footnote{kviswana@sfu.ca}} \\
Department of Physics \\
Simon Fraser University, Burnaby, British Columbia, Canada, V5A 1S6.

\vspace{0.5cm}

Andrew DeBenedictis {\footnote{adebened@sfu.ca}}\\
Department of Physics, \\
and\\
The Pacific Institute for the Mathematical Sciences,\\
Simon Fraser University, Burnaby, British Columbia, Canada, V5A 1S6.
\end{center}

\newpage  
\vspace{0.5cm}

\begin{center}
 {\noindent{\bf{ABSTRACT}}}
\end{center}

\noindent A five-dimensional gravity theory, motivated by the brane-world picture, with  Kaluza scalar 
in the 5 - dimensional metric as 
$g_{55}(r); r=\sqrt{x^2+y^2+z^2}$, is considered near the possible singularity (small distance scales where gravity is strong) 
and is shown to give rise to  a positive contribution to the Raychaudhuri  equation. This inhibits the focusing of world lines and contributes to non - focusing of the worldlines in the 5-dimensional space. It is also shown that 
the results extend to time dependent cases such as those relevant for black hole interiors and cosmology.  

\vspace{0.2cm}

\newpage  

\noindent{\bf{1. Introduction}}

\vspace{0.5cm}

An interesting question in the field of gravitation is whether there is a simple way to avoid the singularities present in certain classical solutions. The 
Standard Cosmological Model based on Einstein's theory of gravitation, for example, implies that the universe began with a 
big bang singularity. The fascinating arena of black holes also possess singularities. Within the framework of Einstein's theory, the above singularities cannot be avoided without imposing exotic matter of some sort. This can be understood, for instance, by the  Raychaudhuri equation \cite{ref:raychaud} which in the absence of torsion, exhibits focusing of geodesics converging to the singularities \cite{ref:hawkel}. In order to avoid such singularities without imposing exotic matter, one has to go beyond Einstein's theory of gravitation. This has been investigated in the brane-world scenario
and the recent studies indicate possible avoidance of the big bang singularity \cite{ref:dvali} - \cite{ref:brax}.

\vspace{0.5cm}

  Studies in string 
theory motivated non - singular cosmologies \cite{ref:nonsing1}, \cite{ref:nonsing2} can avoid the big bang singularity. It has also been shown that in effective loop quantum gravity theories singularities can be avoided \cite{ref:lqc}, \cite{ref:lqbh} . Ellis and Maartens \cite{ref:elmar}, 
and Ellis, Murugan and Tsagas \cite{ref:elmur} proposed the emergent scenario in which the universe stays in a static 
past eternally and then evolves to a subsequent inflationary era, suggesting that the universe originates 
from Einstein static state rather than a big bang singularity. An emergent scenario has been made possible 
in the modified theories of gravity such as $f(R)$ gravity, loop quantum gravity \cite{ref:asht} and in Einstein - Cartan 
theory \cite{ref:ecart}. {\it{These studies motivate the consideration of higher dimensional gravity as a possible candidate 
for avoiding a singularity}}. Further motivation is provided by the trace anomaly of Conformal Field Theory 
dual to a 5 - dimensional Schwarzschild AdS geometry \cite{ref:ads} in which $H^4$ ($H$ being the Hubble parameter) 
terms are present in the equation for $\dot{H}$ and which leads to an infinite age of the universe, avoiding 
the singularity. Similar resolution of the singularity comes from the corrections to Raychaudhuri equation 
in the brane world scenario \cite{ref:branew}, in the approach using the 'generalized uncertainty principle' of quantum 
gravity \cite{ref:lidsey} and in the quantum corrected Friedmann equations \cite{ref:fried1}, \cite{ref:fried2}. An interesting analysis from black hole-brane interactions has been done in \cite{ref:Stojkovic}. 

\vspace{0.5cm}

While a theory of quantum gravity is far from being realized, a quantum corrected 
Raychaudhuri equation has been proposed by Das \cite{ref:das1} and this was the basis to obtain the corrected 
Friedmann equation for $\dot{H}$ by Ali and Das \cite{ref:das2} which avoids the big bang singularity, 
predicting infinite age for the universe. The said corrections to the Raychaudhuri equation cause 
defocussing of the geodesics thereby avoiding the singularity. We find in this manuscript that there is also a simple classical mechanism to produce a similar defocussing term in 5-dimensional world.    

\vspace{0.5cm}

From the above studies, it is clear that in order to avoid the singularity, one needs to modify gravity 
such that defocussing of the geodesics occurs. One way to modify Einstein's theory of gravity is to 
consider five dimensional gravity (without electromagnetic fields) near the singularity (small scales 
where gravity is strong), a minimum modification.  

\vspace{0.5cm}

A direct way to understand the possible avoidance of certain space-like singularities is to consider the Raychaudhuri 
equation. This equation in 4 - dimensional gravity is 
$\frac{d\Theta}{ds}=-\frac{{\Theta}^2}{3}-{\sigma}_{\mu\nu}{\sigma}^{\mu\nu}+{\omega}_{\mu\nu}{\omega}^{\mu\nu}
-R_{\mu\nu}u^{\mu}u^{\nu}+(\dot{u}^{\mu})_{;\mu}$, where $\Theta=u^{\mu}_{;\mu}$ characterizing the volume 
of the collection of particles with 4-velocity $u^{\mu}$ as they fall under gravity. That is, $\Theta$ provides a description of the expansion or contraction of a material body containing streamlines. The quantity ${\sigma}_{\mu\nu}$ is the 
symmetric tensor representing the shear, ${\omega}_{\mu\nu}$ is the antisymmetric tensor representing the 
vorticity and the last term $(\dot{u}^{\mu})_{;\mu}$ vanishes when the particles travel on their geodesics in 4-d theory. 
The vorticity causes expansion while the shear contraction. In the absence of vorticity or exotic matter,  
the geodesics contract or focus causing the universe to have a beginning a finite time ago, creating the big bang singularity or black hole spacelike singularity \cite{ref:hawkel}. One 
could obtain solutions with shear and no vorticity; but not with vorticity and no shear \cite{ref:narlikar} in Einstein's 
theory. The singularity theorems of Penrose and Hawking use this feature to state that there is an inevitable 
spacetime singularity \cite{ref:hawkel}, \cite{ref:narlikar}. In the absence of vorticity and shear, if the last term  $(\dot{u}^{\mu})_{;\mu}$ 
exists and is positive, then defocussing of the world lines occurs thereby softening, or potentially avoiding the singularity. 

\vspace{0.5cm}

Thus, attempts to avoid the singularity require either use of complicated field theoretic models of matter or modified gravity. The quantum corrections to the Raychaudhuri equation in  \cite{ref:branew}, \cite{ref:lidsey}, \cite{ref:das1} achieve this, preventing focusing of geodesics. This feature in cosmological considerations led to the avoiding of the big bang singularity with the universe without a beginning. Since certain black hole interiors have similar causal structure, the arguments may also apply to those singularities.

\vspace{0.5cm}

It is worthwhile to examine whether a non-focusing term of the world lines similar to the above studies could emerge {\it{classically}} in higher dimensional gravity with minimum modifications. 

\vspace{0.5cm}

The aim of this paper is to show a defocussing term of world lines arises by considering five-dimensional Kaluza theory with
fifth dimension at small scales (near the singularity) where gravity is expected to be strong - a minimal modification 
of the 4-d gravity. We hasten to add that we interpret our results modestly, in that we show that a defocussing term appears similar to what is found in the models cited above, but from the consideration of classical general relativity without exotic or quantum matter. The price to pay in this case is the introduction of higher 
dimensions. The defocussing of world lines occurs in the 5-dimensional world. 

\vspace{0.5cm}

In Section 2, we show this by considering static and spherically symmetric 5-d space time. The '55'
part of the 5-d metric, a 4-d scalar, is shown to be responsible for the non-focusing feature. We also extend the analysis to time-dependent domains of the type applicable to some black hole interiors.

\vspace{0.5cm}

\noindent{\bf{2. Effect of Kaluza scalar on Raychaudhuri equation}}

\vspace{0.5cm}

We consider gravity in 5 - d spacetime as in Kaluza-Klein theory with a non-compact or compact fifth dimension. (If the fifth dimension is compact, one will be dealing with specific discrete modes.) Non-compact 
Kaluza-Klein theory has been considered by Wesson \cite{ref:wesson1} and in \cite{ref:wesson2}-\cite{ref:bellin} as Space-Time-Matter theory. The 5 - dimensional metric 
chosen corresponds to (without electromagnetism) 
\begin{eqnarray}
 (dS)^2&=&g_{\mu\nu}\ dx^{\mu}dx^{\nu}-g_{55}(r)\ (dx^5)^2,  \label{eq:linelement}
\end{eqnarray}
where $g_{\mu\nu}$ is the 4-d metric and  $g_{55}=g_{55}(r)\ ;\ r^2=x^2+y^2+z^2$ the Kaluza scalar. The line element 
(\ref{eq:linelement}) corresponds to a 5-d spacetime which can be viewed as 4-d gravity with Kaluza scalar $g_{55}(r)$.  The quantity $dS$ is the full 5-d element in {\it{all}} that follows. That is, it is an invariant under the full 5-d diffeomorphism group. The 5-d spacetime will correspond to a solution of the 5-d vacuum Einstein equations ${\tilde{R}}_{AB}=0;\ A,B=0,1,2,3,5$, 
where ${\tilde{R}}_{AB}$ is the Ricci tensor in the full 5-d spacetime. We
study the strong gravity regime by this 5-d gravity. 

\vspace{0.5cm}

A remarkable consequence of this metric in (\ref{eq:linelement}), is that 5- d world line equation, restricting to 4-d coordinates,
 has an 'acceleration' term from  
 $g_{55}(r)$ (by decomposing $A,B,C$ to $\mu,\nu,\lambda$ and separating out the $A,B,C=5$ components), namely 
\begin{eqnarray}
\frac{d^2x^{\mu}}{dS^2}+{\bigtriangleup}^{\mu}_{\nu\lambda}\ \frac{dx^{\nu}}{dS}\ \frac{dx^{\lambda}}{dS}&=&\frac{1}{2}\ 
\frac{a^2}{g_{55}^2}g^{\mu\lambda}\left({\partial}_{\lambda}g_{55}\right), \label{eq:p2}
\end{eqnarray}
where $a$ is a constant along the world line \cite{ref:flint}, a consequence of the independence of the metric components in (\ref{eq:linelement}) 
on the fifth coordinate $x^5$. This can be seen as: The world line equation in 5-dimensional theory is 
\begin{eqnarray}
\frac{d^2z^A}{dS^2}+{\bigtriangleup}^A_{BC}\ \frac{dz^B}{dS}\ \frac{dz^C}{dS}&=&0, \nonumber 
\end{eqnarray}
where ${\bigtriangleup}$ is the 5-dimensional connection coefficients. The above world line equation can be rewritten as 
\begin{eqnarray}
\frac{d}{dS}\left(g_{AB}\frac{dz^B}{dS}\right)-\frac{1}{2}({\partial}_Ag_{CD})\ \frac{dz^C}{dS}
\ \frac{dz^D}{dS}&=&0. \nonumber
\end{eqnarray}
The $A=5$ part of this equation gives 
\begin{eqnarray}
\frac{d}{dS}\left(g_{5B}\frac{dz^B}{dS}\right)&=&0, \nonumber 
\end{eqnarray}
since the 5-dimensional metric are chosen to be independent of $x^5$ the fifth coordinate. So we have 
\begin{eqnarray}
g_{5B}\frac{dz^B}{dS}&=& a, \nonumber 
\end{eqnarray}
where $a$ is a constant along the 5-d  world line. In our case as $g_{5\mu}=0$ as we are not considering electromagnetic fields and 
so $a=g_{55}\frac{dx^5}{dS}$. The `acceleration' stated above is the acceleration with respect to the full $dS$ in the 
5-dimensional world.  

\vspace{0.5cm}

In order to see how the scheme works in the case of a non-compact fifth dimension, let us consider the 5-d action
\begin{eqnarray}
S_{5}=\frac{1}{16\pi {G}_5}\int d^4x\,dx^5\, \sqrt{G}\,\hat{R}\, \label{eq:action}
\end{eqnarray}
where ${G}_5$ is the 5-d gravitational constant, $G$ is 5-d metric determinant and $\hat{R}$ is the 5-d
Ricci scalar. As none of the components of the metric $G_{AB}$ and hence $\hat{R}$ depend on $x^5$, we see 
the split (4+1) as 
\begin{eqnarray}
S&=&\frac{1}{16\pi G_5}\ \int dx^5\ \int d^4x \sqrt{-G}\hat{R}, \nonumber \\
 &\equiv&\frac{1}{16\pi G_4}\int d^4x\ \sqrt{-G}\hat{R}, \nonumber 
\end{eqnarray}
where we define the effective 4-dimensional gravitational Newton constant $G_4$ as 
\begin{eqnarray}
\frac{1}{G_4}&\equiv&\frac{1}{G_5}\int dx^5. \nonumber 
\end{eqnarray}
Then the fifth coordinate does not appear in the effective action. This leads to 4-dimensional action,
which given the structure of $G_{AB}$ and $\hat{R}$ corresponds in 4-dimensions gravitation with 
scalar $g_{55}(r)$ coupling. The non compact $x^5$ is made small so that $G_4$ remains finite. The fifth 
coordinate due to its small dimensions is not an observable at present day experiments.  
In the resulting expressions we only use ${G}_4$ and so the motion in the fifth coordinate becomes unobservable. 
Of course, one may also pick that the fifth dimension is compact, in which case as long as the pitch (in the 4-d hypersurface) due to the motion in the fifth dimension is small, the motion in the fifth dimension will be unobservable. 

\vspace{0.5cm}

The occurrence of the `acceleration' term (right side in (\ref{eq:p2})) in writing down the 
4-d worldline equation from 5-d Kaluza theory, has been realized earlier Schmutzer \cite{ref:schmutzer}, Kovacs \cite{ref:kovacs}, Gegenberg and 
Kunstatter \cite{ref:gegen}, Mashhoon, Liu and Wesson \cite{ref:mash}, Wesson, Mashhoon, Liu and Sajko \cite{ref:wesmash} and in the brane world scenario 
by Youm \cite{ref:youm}. This author studied the effect of this acceleration to restore causality in brane world \cite{ref:partha1}. 

\vspace{0.5cm}

The Raychaudhuri equation in 5-d spacetime, from 5-d theory, describing the evolution of a collection of particles 
following their world line (\ref{eq:p2}) in the 5 - dimensional world, is given by 
\begin{eqnarray}
\dot{\Theta}&=&-\frac{{\Theta}^2}{4}-2{\sigma}^2+2{\omega}^2-\frac{R^{(4)}}{2}+
({\dot{u}}^{\mu})_{;\mu}\, \label{eq:raychaud}
\end{eqnarray}
 where $u^{\mu}=\frac{dx^{\mu}}{dS}$, $2{\sigma}^2={\sigma}_{\mu\nu}{\sigma}^{\mu\nu}$, $2{\omega}^2={\omega}_{\mu\nu}
{\omega}^{\mu\nu}$, $R^{(4)}$ is the 4-dimensional Ricci scalar.  The subscript $;$ 
stands for covariant derivative (the explicit form of this is given below). For simplicity the cosmological constant $\Lambda$ is set to zero. We have considered 
the 5 - d Raychaudhuri equation from the 5 - d theory, in terms of 4 - d quantities and $dS$ refers to the 5-dimensional
line element.  
This has the factor $\frac{1}{4}$ in front of ${\Theta}^2$ instead of $\frac{1}{3}$. This numeric has very little 
effect in the discussion. In (\ref{eq:raychaud}), the last term involves 
$\dot{u}^{\mu}\ =\ u^{\mu}_{;\nu}u^{\nu}$, the possible `acceleration' (orthogonal to $u^{\mu}$) of the collection of 
particles. In view of (\ref{eq:p2}), this term  exists now. This term is: $\dot{u}^{\mu}
=u^{\mu}_{;\nu}u^{\nu}=\left(\frac{du^{\mu}}{dx^{\nu}}+{\bigtriangleup}^{\mu}_{\lambda\nu}u^{\lambda}\right)u^{\nu}\ =\ 
\frac{du^{\mu}}{dS}+{\bigtriangleup}^{\mu}_{\lambda\nu}u^{\lambda}u^{\nu}\ =\ \frac{d^2x^{\mu}}{dS^2}+{\bigtriangleup}^{\mu}
_{\lambda\nu}\frac{dx^{\lambda}}{dS}\ \frac{dx^{\nu}}{dS}\ =\ \frac{a^2}{2}\frac{1}{g_{55}^2}\ g^{\mu\rho}
({\partial}_{\rho}g_{55})\ =\ -\frac{a^2}{2}g^{\mu\rho}({\partial}_{\rho}\frac{1}{g_{55}})$ using (\ref{eq:p2}). So, the 
last term in (\ref{eq:raychaud}) becomes 
\begin{eqnarray}
\left(\dot{u}^{\mu}\right)_{;\mu}&=&-\frac{a^2}{2}\ g^{\mu\rho} D_{\mu}({\partial}_{\rho}\frac{1}{g_{55}})\, \label{eq:accel}
\end{eqnarray}
where $D_{\mu}$ stands for the covariant derivative, namely $D_{\mu}({\partial}_{\rho}\frac{1}{g_{55}})={\partial}_{\mu}
{\partial}_{\rho}\frac{1}{g_{55}}-{\bigtriangleup}^{\sigma}_{\mu\rho}({\partial}_{\sigma}\frac{1}{g_{55}})$. Thus (\ref{eq:raychaud}) is  
\begin{eqnarray}
\dot{\Theta}&=&-\frac{{\Theta}^2}{4}-2{\sigma}^2+2{\omega}^2-\frac{4\pi G}{c^4}(\rho c^2+3p)-\frac{a^2}{2}
g^{\mu\rho}D_{\mu}({\partial}_{\rho}\frac{1}{g_{55}}). \label{eq:Thetadot}
\end{eqnarray}

\vspace{0.5cm}

In this equation, the presence of the last term will be shown to induce expansion
using  ${\tilde{R}}_{AB}=0$. 

\vspace{0.5cm}
 
When compared with the quantum corrected Raychaudhuri equation of Das \cite{ref:das1}, we see that the role of the quantum 
correction $\frac{{\hbar}^2}{m^2}h^{ab}R_{;a;b}$ of \cite{ref:das1} (the other correction term in \cite{ref:das1} has been set to zero in 
\cite{ref:das1}) is played by $-\frac{a^2}{2}g^{\mu\rho}D_{\mu}({\partial}_{\rho}\frac{1}{g_{55}})$, which is {\it{classical}}
in origin.  

\vspace{0.5cm}

The `acceleration' term $-\frac{a^2}{2}g^{\mu\rho}D_{\mu}({\partial}_{\rho}\frac{1}{g_{55}})$ can possibly induce 
expansion in the 5-dimensional space  
if this term turns out to be positive. We examine this for a static spherically symmetric ansatz for 
the 5 - d metric in (\ref{eq:linelement}) as  
\begin{eqnarray}
(dS)^2&=&e^{\mu}c^2(dt)^2-e^{\nu}(dr)^2-r^2\{(d\theta)^2+{\sin}^2{\theta}(d\phi)^2\}-\psi(r)(dx^5)^2, \label{eq:sphereline}
\end{eqnarray} 
where $\mu, \nu$ are functions of $r=\sqrt{x^2+y^2+z^2}$ only and $\psi(r)=g_{55}(r)$ for ease of notation. In (\ref{eq:sphereline}), 
$\mu(r), \nu(r)$ 
and $\psi(r)$ are unknown functions of $r$ to be determined by the 5 - d Einstein vacuum (Ricci flat) equations,  
$\tilde{R}_{AB}=0$, $A,B=0,1,2,3,5$. The choice of static ansatz (\ref{eq:sphereline}) is partially motivated by [9] (proposing 
an emergent scenario in which the universe stays in a static past eternally - universe originating from Einstein 
static state). 

\vspace{0.5cm}

For the metric in (\ref{eq:sphereline}), $\tilde{R}_{tt}, \tilde{R}_{rr},\tilde{R}_{\theta\theta}, \tilde{R}_{\phi\phi}, 
\tilde{R}_{55}$ can be evaluated and the results are given in the Appendix. Now, we show that the additional term in (\ref{eq:accel}) 
due to 'acceleration' from $g_{55}(r)$ is always positive, that is 

\vspace{0.5cm}

{\it{The additional (acceleration) term in (\ref{eq:accel}), namely $-\frac{a^2}{2}g^{\mu\rho}D_{\mu}({\partial}_{\rho}
\frac{1}{g_{55}})$, is always positive for a spherically symmetric general solution with ${\tilde{R}}_{AB}=0$}}. 

\vspace{0.5cm}

Expanding the covariant derivative $D_{\mu}$ and setting $g_{55}(r)=-\psi(r)$ (for ease of notation), we have 
\begin{eqnarray}
-\frac{a^2}{2}g^{\mu\rho}D_{\mu}({\partial}_{\rho}\frac{1}{g_{55}})&=&-\frac{a^2}{2}g^{\mu\rho}{\partial}_{\mu}(
{\partial}_{\rho}\frac{1}{g_{55}})+\frac{a^2}{2}g^{\mu\rho}\ {\bigtriangleup}^{\lambda}_{\mu\rho}({\partial}_{\lambda}
\frac{1}{g_{55}}), \nonumber \\
&=&-\frac{a^2}{{\psi}^3}g^{\mu\rho}{\partial}_{\mu}\psi \ {\partial}_{\rho}\psi+\frac{a^2}{2{\psi}^2}g^{\mu\rho}
{\partial}_{\mu}{\partial}_{\rho}\psi-\frac{a^2}{2{\psi}^2}g^{\mu\rho}{\bigtriangleup}^{\lambda}_{\mu\rho}({\partial}_{\lambda}
\psi). \nonumber \\
& & \label{eq:accelb}
\end{eqnarray}

Since $\psi=\psi(r)$, we have the above expression as 
\begin{eqnarray}
-\frac{a^2}{2}g^{\mu\rho}D_{\mu}({\partial}_{\rho}\frac{1}{g_{55}})&=&-\frac{a^2}{{\psi}^3}g^{rr}(\psi')^2+
\frac{a^2}{2{\psi}^2}g^{rr}\psi''-\frac{a^2}{2{\psi}^2}g^{\mu\rho}{\bigtriangleup}^r_{\mu\rho}\psi', \label{eq:accelc}
\end{eqnarray}
where $\psi'=\frac{d\psi}{dr}\ ;\ \psi''=\frac{d^2\psi}{dr^2}$. For the line element (metric) in (\ref{eq:sphereline}), $g^{rr}
=-e^{-\nu}$ and the connection coefficients $\bigtriangleup$ for (\ref{eq:Thetadot}), via (\ref{eq:accelb}) and (\ref{eq:accelc}), are evaluated in the Appendix. Then we have, 
using them 
\begin{eqnarray}
g^{\mu\rho}{\bigtriangleup}^r_{\mu\rho}&=&\frac{1}{2}e^{-\nu}\mu'-\frac{1}{2}e^{-\nu}\nu'+\frac{2}{r}e^{-\nu}, \label{eq:connection}
\end{eqnarray}
where $\mu'=\frac{d\mu}{dr}\ ;\ \nu'=\frac{d\nu}{dr}$. Substituting (\ref{eq:connection}) in (\ref{eq:accelc}), we find 
\begin{eqnarray}
-\frac{a^2}{2}g^{\mu\rho}D_{\mu}({\partial}_{\rho}\frac{1}{g_{55}})&=&\frac{a^2}{{\psi}^2}e^{-\nu}\{
\frac{(\psi')^2}{\psi}-\frac{\psi''}{2}-\frac{\mu'\psi'}{4}+\frac{\nu'\psi'}{4}-\frac{\psi'}{r}\}. \label{eq:acceld}
\end{eqnarray}

The equation ${\tilde{R}}_{55}=0$ in the Appendix, gives 
\begin{eqnarray}
-\frac{\psi''}{2}-\frac{\mu'\psi'}{4}+\frac{\nu'\psi'}{4}-\frac{\psi'}{r}&=&-\frac{(\psi')^2}{4\psi}, \nonumber 
\end{eqnarray}
and substituting this in $\{....\}$ of (\ref{eq:acceld}), we obtain 
\begin{eqnarray}
-\frac{a^2}{2}g^{\mu\rho}D_{\mu}({\partial}_{\rho}\frac{1}{g_{55}})&=&\frac{3a^2}{4}\ e^{-\nu}\ \frac{(\psi')^2}
{{\psi}^3}.
\end{eqnarray}
Since $\psi(r)>0$ (so as to preserve the sign convention for the metric in (\ref{eq:sphereline})) and $e^{-\nu}$ is always positive,
it follows that 
\begin{eqnarray}
\frac{3a^2}{4}\ e^{-\nu}\ \frac{(\psi')^2}{{\psi}^3} &>& 0. 
\end{eqnarray}
The above result proves that 
\begin{eqnarray}
-\frac{a^2}{2}g^{\mu\rho}D_{\mu}({\partial}_{\rho}\frac{1}{g_{55}})&>& 0, 
\end{eqnarray}
for a general spherically symmetric situation.  Thus, the additional term in the Raychaudhuri equation is always positive for a general spherically symmetric 
situation and therefore is effectively repulsive in the 4-d spacetime, as the world lines tend to defocus. Further, 
the defocussing can be seen also from the following considerations. The effective 4-dimensional action after using 
$G_{AB}$ and $\hat{R}$ becomes explicitly
\begin{eqnarray}
S&=&\frac{1}{16\pi G_4}\int d^4x \sqrt{-g}\sqrt{g_{55}}\left(R-\frac{1}{g_{55}(r)}g^{\mu\nu}({\partial}_{\nu}
g_{55})_{;\ \mu}\right),
\end{eqnarray}
where $R$ is 4-dimensional Ricci scalar and the second term is the kinetic energy term for $g_{55}$. This has 
wrong sign and hence eventually leads to defocussing.  

\vspace{0.5cm}

With the presence of the last term in (\ref{eq:accel}) whose contribution has been shown to be positive in general, the focusing 
of the world lines in 5 - dimensions is inhibited and could cause non-focusing of world lines in 5-dimensional world. 
Before proceeding to cosmological 
implications, the 'geodesic deviation equation' gets modified by the acceleration term in (\ref{eq:p2}) as 
\begin{eqnarray}
\frac{{\cal{D}}^2(\delta x^{\mu})}{{\cal{D}} s^2}&=&-R^{\mu}_{\ \lambda\nu\rho}\ \frac{dx^{\lambda}}{ds}
\ \delta x^{\nu}\ \frac{dx^{\rho}}{ds}+f^{\mu}_{;\nu}\ \delta x^{\nu},
\end{eqnarray}
where $f^{\mu}_{;\nu}=-\frac{a^2}{2}g^{\mu\lambda}D_{\nu}({\partial}_{\lambda}\frac{1}{g_{55}})$, showing the 
deviation $\delta x^{\mu} \neq 0$. The observed deviation acceleration would remain finite. When $\delta x^{\mu}
\neq 0$, the world lines are inhibited to converge, supporting the defocussing feature from the considerations 
of Raychaudhuri equation. 

\vspace{0.5cm}

We can similarly consider the analogous spherically symmetric time-dependent domain (also sometimes known as the T-domain) by extending the study to the following line element:
\begin{eqnarray}
 dS^{2}= e^{\mu(t)}\,(dt)^{2} -e^{\nu(t)}\,(dy)^{2} -t^{2}\left\{(d\theta)^{2} +\sin^{2}\theta\,(d\phi)^{2}\right\} -\psi(t)\,(dx^{5})^{2},
\end{eqnarray}
The above line element is relevant, for example, inside the event horizon of a spherically symmetric black hole where the timelike and 
radial spacelike coordinates change character. The calculations here are very similar to the ones presented for the metric (\ref{eq:sphereline}) so we do not reproduce the calculations in detail. In brief, the five dimensional worldline equation for 4-dimensional $x^{\alpha}$  
in this scenario yields a similar expression as before
\begin{eqnarray}
 \frac{d^{2}x^{\alpha}}{dS^{2}}+{\bigtriangleup}^{\alpha}_{\beta\gamma} \frac{dx^{\beta}}{dS}\frac{dx^{\gamma}}{dS} =\frac{1}{2}\left(\partial_{\beta}g_{55}\right)g^{\alpha\beta}\frac{a^{2}}{(g_{55})^{2}},
\end{eqnarray}
where the constant $a$ is defined as in the previous section. Note though that different metric functions from before will appear from the index sum due to the time dependence of the metric. Also, an analogous result to the expression just before equation (\ref{eq:accel}) is also calculated:
\begin{equation}
 u^{\alpha}_{\;;\beta}u^{\beta}=-\frac{a^{2}}{2}g^{\alpha\rho}\left(\partial_{\rho} \frac{1}{g_{55}}\right).
\end{equation}
This leads to the investigation of the positivity (following a proof similar to the previous one) of the following:
\begin{eqnarray}
 -\frac{a^{2}}{2}g^{\alpha\beta}D_{\beta}\left(\partial_{\rho} \frac{1}{g_{55}}\right)= \frac{3}{4} \frac{a^{2}\dot{\psi}^{2}}{\psi^{3}} e^{-\mu(t)},
\end{eqnarray}
which is indeed positive since $\psi > 0$ due to the Lorentzian structure demanded of the metric. (Here the over-dot represents $\partial_{t}$.)

\vspace{0.5cm}

\noindent{\bf{3.Concluding remarks}}

\vspace{0.5cm}

We have shown that five dimensional world line defocussing may be achieved by considering five dimensional spacetime with 
the '55' component of $G_{AB}$ a Kaluza scalar. 
Qualitatively this is similar to the defocussing which occurs when considering some effective quantum corrected models. 
This shows that such a defocussing effect can also have a purely classical origin which is completely in five dimensional general relativity.  

\vspace{0.5cm}

\noindent {\bf{Acknowledgements}}

\vspace{0.4cm}

R. Parthasarathy is grateful for productive discussions with G. Rajasekaran, R. Rajaraman, A.P. Balachandran and Govind Krishnaswamy.

\vspace{0.5cm}
\newpage

{\noindent{\bf{Appendix}}}

\vspace{0.5cm}

For the metric in (\ref{eq:sphereline}), the non - vanishing five dimensional Christoffel connections are:
\begin{eqnarray}
{\bigtriangleup}^t_{rt}\ =\ \frac{\mu'}{2}; &{\bigtriangleup}^r_{rr}\ =\ \frac{\nu'}{2};& {\bigtriangleup}^r_{tt}\ =\ 
\frac{1}{2}e^{(\mu-\nu)}c^2\mu';\nonumber \\
{\bigtriangleup}^r_{\theta\theta}\ =\ -e^{-\nu}r;&{\bigtriangleup}^r_{\phi\phi}\ =\ -e^{-\nu}r{\sin}^2{\theta};& 
{\bigtriangleup}^r_{55}\ =\ -\frac{1}{2}e^{-\nu}\psi'; \nonumber \\
{\bigtriangleup}^{\theta}_{r\theta}\ =\ \frac{1}{r};&{\bigtriangleup}^{\theta}_{\phi\phi}\ =\ -\sin{\theta}\cos{\theta};&
{\bigtriangleup}^{\phi}_{r\phi}\ =\ \frac{1}{r}; \nonumber \\
&{\bigtriangleup}^{\phi}_{\theta\phi}\ =\ \cot{\theta} \ ;\ {\bigtriangleup}^5_{r5}\ =\ \frac{\psi'}{2\psi},& 
\nonumber \hspace{3.0cm} (A1)
\end{eqnarray}
where the prime $'$ stands for differentiation with respect to $r$. The Ricci tensor
\begin{eqnarray}
{\tilde{R}}_{AB}&=&{\partial}_C{\bigtriangleup}^C_{AB}-{\partial}_B{\bigtriangleup}^C_{AC}+{\bigtriangleup}^C_{DC}
{\bigtriangleup}^D_{AB}-{\bigtriangleup}^C_{DB}{\bigtriangleup}^D_{CA},\nonumber 
\end{eqnarray}
gives the components as
\begin{eqnarray}
{\tilde{R}}_{tt}&=&\frac{1}{2}e^{(\mu-\nu)}c^2\{\mu^{\prime\prime}-\frac{\mu'\nu'}{2}+\frac{{\mu'}^2}{2}+\frac{2\mu'}{r}
+\frac{\mu'\psi'}{2\psi}\}, \nonumber \\
{\tilde{R}}_{rr}&=&-\frac{\mu^{\prime\prime}}{2}+\frac{\mu'\nu'}{4}+\frac{\nu'}{r}-\frac{{\mu'}^2}{4}-\frac{\psi"}{2\psi}
+\frac{\nu'\psi'}{4\psi}+\frac{{\psi'}^2}{4{\psi}^2}, \nonumber \\
{\tilde{R}}_{\theta\theta}&=&1-e^{-\nu}-re^{-\nu}\{\frac{\mu'}{2}-\frac{\nu'}{2}+\frac{\psi'}{2\psi}\}, \nonumber \\
{\tilde{R}}_{\phi\phi}&=&{\sin}^2{\theta}\ {\tilde{R}}_{\theta\theta}, \nonumber \\
{\tilde{R}}_{55}&=&e^{-\nu}\{-\frac{\psi^{\prime\prime}}{2}-\frac{\mu'\psi'}{4}+\frac{\nu'\psi'}{4}-\frac{\psi'}{r}
+\frac{{\psi'}^2}{4\psi}\}.\nonumber \\
& & \nonumber \hspace{10.0cm} (A2)
\end{eqnarray}

\newpage

\vspace{0.5cm}

\bibliographystyle{unsrt}

\end{document}